\documentclass[aps,pre,preprint,groupedaddress]{article}
\usepackage{epsfig}
\usepackage{rotating}
\usepackage{color}
\usepackage{longtable}
\usepackage{multicol}
\usepackage[normalem]{ulem}
\usepackage{hyperref}
\usepackage{breakurl}
\usepackage{graphicx}
\usepackage{authblk}

\begin{document}

\title{Population patterns in World's administrative units}


\author[1]{Oscar Fontanelli}
\author[1]{Pedro Miramontes}
\author[2]{Germinal Cocho}
\author[3]{Wentian Li}
\affil[1]{Departamento de Matem\'{a}ticas, Facultad de Ciencias,
Universidad Nacional Aut\'{o}noma de M\'{e}xico,
Mexico City, M\'{e}xico}
\affil[2]{Instituto de F\'{i}sica, Universidad Nacional Aut\'{o}noma
de M\'{e}xico,
Mexico City, M\'{e}xico}
\affil[3]{The Robert S. Boas Center for Genomics and Human Genetics,
The Feinstein Institute for Medical Research,
Northwell Health, Manhasset, NY,
USA.}

\maketitle

\begin{abstract}

While there has been an extended discussion concerning city population distribution, little has been said about administrative units. Even though there might be a correspondence between cities and administrative divisions, they are conceptually different entities and the correspondence breaks as artificial divisions form and evolve. In this work we investigate the \-po\-pu\-la\-tion distribution of second level administrative units for 150 countries and propose the Discrete Generalized Beta Distribution (DGBD) rank-size function to describe the data. After testing the goodness of fit of this two parameter function against power law, which is the most common model for city population, DGBD is a good statistical model for $73\%$ of our data sets and better than power law in almost every case. Particularly, DGBD is better than power law for fitting country population data. The fitted parameters of this function allow us to \-cons\-truct a phenomenological characterization of countries according to the way in which people are distributed inside them. We present a computational model to simulate the formation of \-ad\-mi\-nis\-tra\-tive divisions and give numerical evidence that DGBD arises from it. This model along with the DGBD function prove adequate to reproduce and describe local unit evolution and its effect on population distribution.
 
\end{abstract}

\section*{Introduction}

\noindent The inhabitable area of the world is divided into political distinct units which may be countries, dependent territories, et cetera. For the purpose of internal government and management, these units are often subdivided into smaller areas called administrative divisions or units. For example, the USA is divided into states (first level division), which in turn are divided into counties (second level division); China is divided into provinces blue and direct-controlled municipalities (first level), which are split into prefectures (second level), which are divided into counties (third level) and so on. Among other things, local administrative units can be seen as  socially constructed entities, which serve as spacial scenarios of social and economic processes \cite{lopez}.\\

\noindent The heterogeneity of administrative units between and inside countries is wide: some countries, like Andorra and Cyprus, have only first level divisions (parishes and districts respectively). Most of them have divisions up to the second level, while some others, France and Italy for example, have divisions up to the third level (arrondissements and comunes). Furthermore, the structure and size of administrative units is a hallmark of a country's internal organization. For instance, China reorganized its administrative unit system in view of the economic reforms in the decade of 1970, yielding economic benefits for the central cities \cite{laurence}. In contrast, many developing countries have attempted to establish more decentralized local governments by increasing the number of administration divisions \cite{grossman}. These examples also illustrate the fact that the territorial organization within countries is in constant evolution. Even in the absence of large administrative and political reforms, administrative units are constantly being created, destroyed, merged or split \cite{gantner}. The high degree of diversity and complexity in the internal territorial administration of countries worldwide relies partly in the population distribution over them \cite{christenson}. \\

\noindent A crucial factor for life within a given territory is its population, which does not distribute randomly over the available space  \cite{eeckhout}. For example, there are studies that show that a city's \-po\-pu\-la\-tion strongly correlates with many of the features of city's inhabitants: mean income, number of registered patents per capita, \-cri\-mi\-na\-li\-ty rates, land value and rent prices \cite{bettencourt07,bettencourt10}. In this context the study and understanding of the geographical distribution of population within a given country or region becomes relevant, since it is a \-ne\-ce\-ssa\-ry step for the development of theories that could accurately describe the evolution of human agglomerations \cite{batty08,batty13}. \\

\noindent Almost all studies on population distribution focus on city population (see for instance \cite{soo,rozenfeld08,holmes,jiang,rozenfeld,courtat,batty-masucci,masucci}, as well as the extended debate on whether city population follows a power law or a lognormal distribution \cite{levy09,eeckhout09,malevergne} and whether Gibrat's law holds for city growth \cite{gabaix,saichev,nassar}). However, the literature on administrative unit population distribution is scarce. Some examples were given in \cite{beta-plos1}, where the authors analyze population distribution for administrative divisions in China, Mexico and Spain. In the present work we carry out a much more comprehensive study. Does Zipf's law hold for administrative unit population as it does for big cities? Are there any traces of an ubiquitous behavior? What is the effect on population distribution of external agents delineating artificial boundaries for \-ad\-mi\-nis\-tra\-tive territories?  In this paper we seek to respond to these questions by addressing two main issues: 1) provide a description and characterization of population distribution for secondary administrative divisions (SAU's from now on) for a set of 150 countries; in particular, we discuss the validity of power laws and propose to apply a two parameter rank-size function to fit the data; 2) propose a computational model to describe the process of administrative unit formation and development, from which our two parameter representation arises. This model simulates the evolution of a population distribution in a territory subject to constant spatial divisions made by local authorities. \\

\noindent The rank-size function that we propose to apply is the two parameter Discrete Generalized Beta Distribution (DGBD), which has been used to adjust rank-size and rank-frequency observations in natural, social and even artistic \-phe\-no\-me\-na \cite{beta-joi,beta-mus,beta-plos1,beta-entropy,beta-jql1,beta-pra,beta-com,beta-jql2,beta-ausloos}. In particular, it has been used to fit the size distribution of natural cities \cite{X-Li}. Among the processes that have been proposed to explain its apparent ubiquity we mention multinomial processes \cite{naumis}, restricted subtraction of random variables \cite{beltran} and birth-death processes \cite{roberto}. In order to model the formation of SAU's we implement a version of the \emph{split-merge process}, a computational mechanism in which administrative units are created and destroyed by joining and dividing them, emulating the role of governments delineating internal boundaries. This process was originally proposed in \cite{splitmerge}. \\

\noindent The structure of the paper is the following: first we review the DGBD function and derive some of its relationships it with some common probability distributions. Next we analyze the population distribution in the population of SAU's for 150 countries, evaluate the goodness of fit of DGBD and compare its performance with power laws.  Then we take the samples where DGBD is a better model than power law and use the fitted DGBD parameters to construct a characterization of countries according to the way in which people internally spread. After that we use the split-merge process to simulate SAU's formation and give numerical evidence that it leads to population distributions well described by DGBD. Finally, we discuss how the differences between cities and administrative units naturally motivate the split-merge process, its  suitability for outlining the administrator's role in defining arbitrary boundaries and the appropriateness of DGBD in representing these phenomena. 

\section*{The rank-size representation and the DGBD function}

It is said that a random phenomenon presents a heavy tail when there is a relatively high probability for large, rare events. When this happens, it is customary to describe it via the rank-size or rank-frequency distribution, instead of the probability density function (pdf) \cite{sornette}. This is often the case for population distributions where a small amount of high populated regions encompass the majority of the population. Thus, we will adopt the rank-size representation to describe the population distribution within a country.  This representation does not require binning, avoiding the problem that bins are under-sampled at the tail.  One way to link these two representations is defining the rank of a certain observation proportional to the probability of making a larger observation.  More precisely, let $X$ be a continuous random variable with density $f(x)$ and $\{x_i\}_{i=1}^n$ a collection of $n$ realizations. Let $x_M$ and $x_m$ be the maximum and minimum of the latter. We define the \emph{rank} of $x$, denoted by $r(x)$, as

\begin{equation}
\label{rank}
\displaystyle r(x)=r_m+(r_M-r_m)\int_x^{x_M}f(t)dt,
\end{equation}

\noindent where $r_M=r(x_m)$ and $r_m=r(x_M).$ The DGBD rank-size or rank-frequency function is a two parameter function that outperforms other common distributions in describing scaling behaviors over a large sample of phenomena \cite{beta-plos1}. It is defined by

\begin{equation}
\label{DGBD}
f(r)=\frac{C(r_M+r_m-r)^b}{r^a},
\end{equation}

\noindent where $r_M$ and $r_m$ denote maximum and minimum rank (usually $r_m=1$ and $r_M$ equals the number of observations), $b$ and $a$ are parameters to be estimated and $C$ is a \-nor\-ma\-li\-za\-tion constant. Now we will derive some relationships between DGBD and common probability functions. First we observe that DGBD reduces to power law when $b=0$. In particular, it is Zipf's law when $b=0$ and $a=1$. In second place, suppose $X$ is a random variable uniformly distributed over the interval $(\alpha, \beta)$. This means that $X$ has the pdf $f_X(x)=\frac{1}{\beta-\alpha}1_{(\alpha, \beta)}$. Thus, for $x$ in $(\alpha, \beta)$, Eq.(\ref{rank}) implies that
$$
r=1+(r_M-1)\frac{\beta -x}{\beta-\alpha}.
$$

\noindent Solving this last equation for $x$ we get after some arrangements
$$
x=\frac{\beta-\alpha}{r_M-1}\left(\frac{(r_M-1)\beta}{\beta-\alpha}+1-r\right),
$$

\noindent which reduces to DGBD with $b=1$ and $a=0$ by taking $\beta=\frac{r_M}{r_M-1}$ and $\alpha=\frac{1}{r_M-1}$ on the limit when the maximum rank (consequently the number of observations) $r_M$ tends to infinity. Thirdly, we observe that for a zero variance distribution, located at $x_0$ and with a delta pdf function $f_X(x)=\delta (x-x_0)$, Eq.(\ref{rank}) reduces to $r=r_M$, which is a DGBD with $b=a=0$ and constant of normalization $C=r_M$. Lastly, it has been recently shown that when $b=a$ the pdf associated to the DGBD can be analytically derived, yielding a novel probability law called Lavalette distribution, which closely resembles the lognormal distribution \cite{fontanelli}.  In summary,

\begin{enumerate}
\item A power law is represented by a DGBD with $b=0$. If additionally $a=1$, it is Zipf's law.
\item A uniform distribution is represented by a DGBD with $b=1$ and $a=0$ on the limit when the number of observations is very large.
\item A point located distribution is represented by a DGBD with $b=a=0$.
\item When $a=b$, DGBD represents a random phenomenon with a distribution approximately equal to lognormal. 
\end{enumerate}

\noindent Thus, DGBD can be used to represent the population distribution in a completely ``flat'' country, where all cities or administrative units have the same population ($b=a=0$) or, at the other extreme, in a completely disordered country where population is distributed fully at random (uniform distribution, $b=1$ and $a=0$).  

\section*{DGBD and administrative unit population}

\begin{figure}[!h]
\epsfig{file=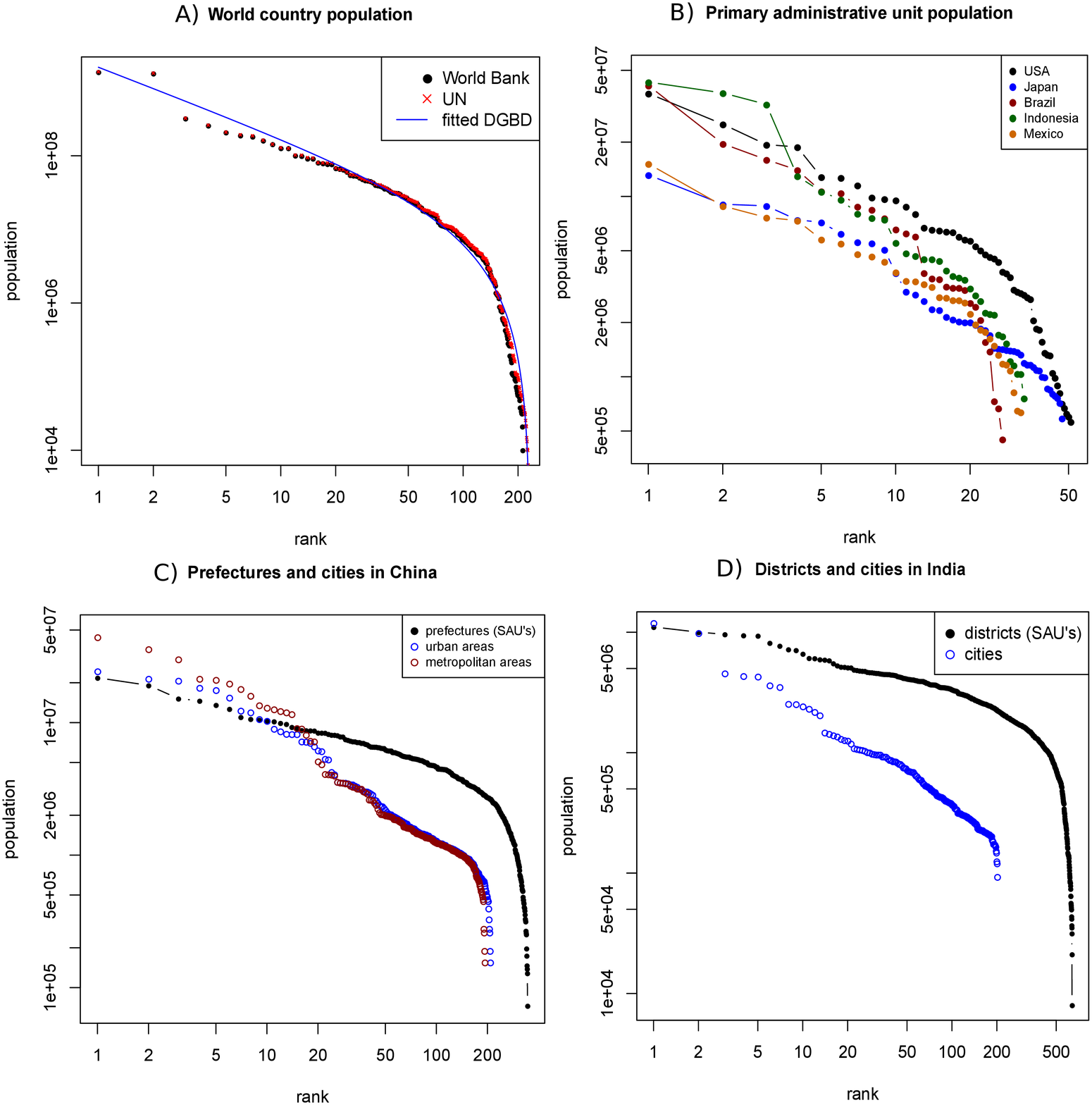,width=0.6\linewidth}
\caption{{\bf Examples of deviations from power laws}. A) Ranked population of countries and territories of the world according to data of the World Bank and the UN, red line is the fitted DGBD to the UN data; B) ranked population of primary administrative units in five different countries (states in the US, Brazil and Mexico, prefectures in Japan and provinces in Indonesia); C) ranked population of prefectures (second level administrative division) urban and metropolitan areas; D) ranked population of districts (second level administrative units) and cities in India. All plots are in log-log scale. }
\label{china-india}
\end{figure}

\noindent While city population in different countries often exhibit power law or lognormal behaviors, when artificial divisions come into consideration it is unclear what the distribution should be. To introduce the subject, we present on Fig.(\ref{china-india}) several examples of ranked population of administrative units. First there is population of all countries and territories of the world, which we can think as zero-level administrative units; as a matter of example of primary or first level administrative units (PAU from now on) we present ranked population of these units for five different countries; because number of PAU's is low for many countries, we will focus most of our analysis on secondary administrative units; the third panel shows ranked population for prefectures (secondary administrative units) in China, together with city population according to two different definitions: urban and metropolitan areas (information and data of these taken from \url{https://en.wikipedia.org/wiki/List_of_cities_in_China_by_population_and_built-up_area}); finally, we show ranked population of cities (\url{https://en.wikipedia.org/wikiList_of_cities_in_India_by_population}) and districts (SAU's) in India. Normally, power laws are expected when the rank-size or rank-frequency plot is approximately a straight line in the log-log representation. Although this is by no means a statistical evidence to claim that a data set or phenomenon is a power law, it is a necessary condition that potential power law candidates must fulfill. Deviations from power law behavior can be appreciated in Fig.(\ref{china-india}). For instance, there is a breakdown in the distribution tail of country population, yet DGBD provides a good fit at both tails, as the blue line shows; these deviations are more difficult to appreciate in PAU population, but are more evident for SAU's, as panels C) and D) show. Regarding city population, we see in our examples that deviations from power law appear in the low-population regime, but  it is a good model for the largest cities.\\

\noindent We obtained SAU population data from the database Statoids (\url{http://www.statoids.com}, last consulted on June 2016, detailed information on sources and dates of each data set can be found in this site), which gave us the SAU population for 150 countries. From these, we take into account only those where the number of SAU's is greater than $10$, so we have a minimum number of points to perform a regression analysis. For each country we ordered its SAU population by rank and fitted the data to a DGBD via a linear regression of the logarithmic transform of Eq.(\ref{DGBD}). As we show in the Appendix, the coefficients of determination $R^2$ are invariably high, which is a first sign that DGBD may be indeed a good fit. However, the coefficient of determination is not enough statistical evidence in favour of DGBD as a good statistical model. To test the goodness of fit of the DGBD we follow a procedure similar to that proposed in \cite{clauset}. First we compute the Kolmogorov-Smirnov statistic to compare the data with a theoretical DGBD. Next we simulate a large number of DGBD data with the fitted parameters and compute the K-S statistic of each simulated sample. The K-S statistic is a measure of distance between the data and the reference distribution. We compute the fraction of simulated samples that are farther from DGBD than our population data; this fraction gives an estimate of the p-value of the DGBD hypothesis. A larger p-value means random chances mostly lead to a worse fitting, so our fitted function is not bad enough to be rejected. On the other hand, a small p-value implies that our fitting performance is on the worse end among random chances, thus not good enough. With this method, the specific value of the estimation depends on some specific choices, for example how to manage samples with the same K-S distance, the number of replicates, etc. In particular, two countries with different number of SAU's may not be comparable in their p-values, because there is a tendency of the empirical p-value to be higher with a low maximum rank of the data. Nevertheless, a large p-value still indicates the fitted model can not be rejected, after the removal of countries with a very low number of SAU's.   \\

\begin{figure}[!h]
\epsfig{file=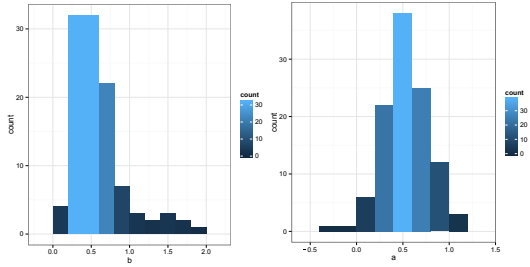,width=0.6\linewidth}
\caption{{\bf Histograms of fitted $b$ and $a$ for the 108 selected countries}. The parameters $b$ and $a$ were estimated by a linear regression of the logarithm of the DGBD. 
}
\label{fig2}
\end{figure}

\noindent We take the countries for which the p-value is  larger than 0.05 (not enough evidence to reject DGBD). We would like to compare the DGBD model for these countries with a power law, which constitutes the traditional model for city population. We do this with Akaike information criterion (AIC), that measures the relative quality of a statistical model. The model that exhibits a lower $AIC=2k+n\log \left( \frac{RSS}{n}\right)$ is the better model ($k$ is the number of estimated parameters, $n$ the sample size and $RSS$ the residual sum of squares). Even though power law is a particular case of DGBD, this criterion takes into account the number of parameters of each model, giving a way to discern between a DGBD model with $b=0$ and a pure power law. We take only those countries where $AIC_{\mbox{\small{DGBD}}}$ is less than $AIC_{\mbox{\small{power law}}}$. Column named AIC on table \ref{allcountries} in the appendix gives the difference $\log(AIC_{\mbox{\small{DGBD}}})-\log(AIC_{\mbox{\small{power law}}})$. Thus, we discard all countries that have not enough SAU's to do a regression analysis, the countries for which the K-S test rejects the DGBD and those for which power law has a lower Akaike information criterion than DGBD. There are countries where DGBD performs better than power law according to AIC, but it is rejected as a consequence of a small p-value. For our study, we looked for countries that satisfied both criteria. Ranked population and fitted DGBD for each country that we studied can be seen in Fig.(\ref{150plots}) in the appendix. \\

\noindent After these three selection criteria a set of 108 countries remains. Mean and standard deviation of the sample size are $176.9$ and $251.2$ respectively. The mean and standard deviation for the fitted $b$ parameter are $0.58$ and $0.35$, while the mean and standard deviation for $a$ parameter are $0.53$ and $0.25$. We show in Fig.(\ref{fig2}) histograms of $b$ and $a$, which indicate that they are not randomly distributed, but rather clustered around central values. There are two cases, Slovenia and Virgin Island US, for which $a<0$. When this happens, Eq.(\ref{DGBD}) fails to represent a rank-size distribution because it is not monotonous; however, the fittings are still good; usually this means that the maximum of the fitted curved is reached below $rank=1$ \cite{beta-mus}. \\

\begin{figure}[!h]
\epsfig{file=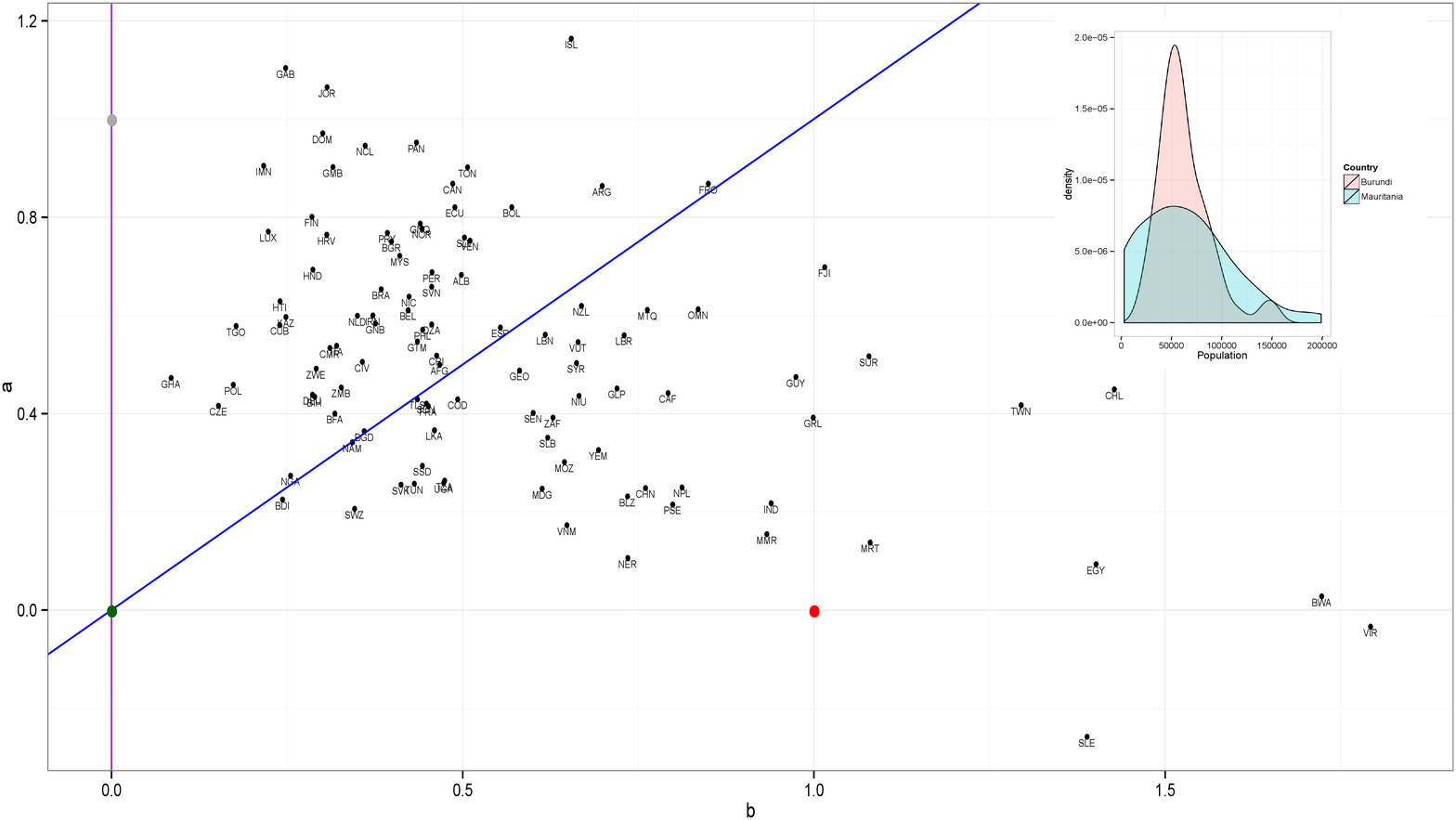,width=0.9\linewidth}
\caption{{\bf Dispersion plot of fitted $b$ and $a$ for the 108 selected countries.} Each country is indicated by its three letter ISO3 code. In purple the line $b=0$, which represents a perfect power law; in blue the line $b=a$, which is the Lavalette rank-size function; gray point is $(0,1)$ a pure Zipf's law; green is $(0,0)$, representing a point located probability distribution; red point is $(1,0)$, probability uniform distribution. We also show the density histogram for two countries: Mauritania, whose parameters indicate a closeness to a uniform distribution, and Burundi, whose parameters suggest a distribution similar to a very narrow lognormal.}
\label{fig3}
\end{figure}

\noindent The fitted DGBD $b$ and $a$ parameters for these countries are shown in the dispersion plot Fig.(\ref{fig3}). Countries are indicated by their three letter ISO3 code. The gray, green and red dots represent idealized regions following perfect Zipf's law, delta and uniform distributions respectively. We indicate with purple the vertical line $b=0$, representing perfect power laws and with blue the line $b=a$, representing the Lavalette distribution which, as we mentioned, has a close resemblance with the lognormal distribution. There has been debate about whether lognormal distribution is a better representation for city population than power laws. It might be said that for SAU population, countries with $b<a$ (between the blue and the purple lines on the diagram) are somewhere between these two distributions, $b<<a$ indicates that power law fits the data well while countries below the blue line ask for a different model. \\

\noindent The dispersion diagram allows a quantitative measure of the distance between the SAU population distribution within a country and an idealized Zipf's law, a completely ordered or disordered population distribution, etc. We propose the euclidean distance on the $<b,a>$ plane to measure this. For example, Myanmar and Mauritania are close to being disordered (uniform distribution) while Burundi and Nigeria are close to being delta distributed (and also close to Lavalette, implying that they have very narrow lognormal-like distributions). These observations are confirmed by observing the histogram densities of Burundi and Mauritania, shown in Fig.(\ref{fig3}). Indeed, we see that Mauritania has a wide pdf, somewhat close to the constant pdf of a uniform random variable, while Burundi exhibits a taller and narrower pdf. \\

\begin{figure}[!h]
\epsfig{file=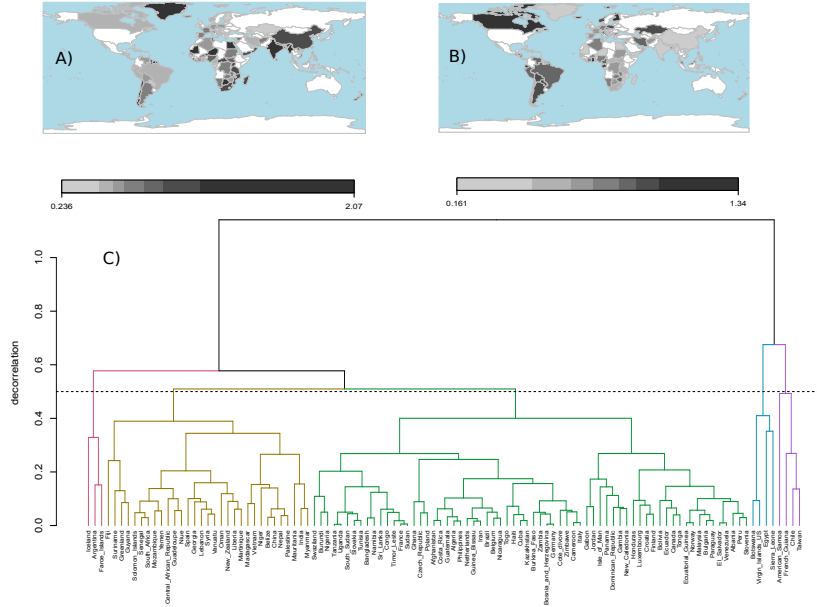,width=0.9\linewidth}
\caption{A) World map indicating distance in the parameter space to a country with a uniform distribution, B) distance to a perfect Zipf's law, C) dendrogram showing five families of kindred countries according to its internal population patterns.}

\label{maps}
\end{figure}

\noindent Even if the internal partition of a country is determined by a central administration, the SAU system might not be completely arbitrary, there may be climatic and geographical factors constricting internal divisions and subdivisions. To investigate this issue we present in Fig.(\ref{maps}) world maps with the euclidean distance of each country to a Zipf's law and to a country with a uniform probability distribution. Countries and regions shown in white are those for which DGBD is not a good statistical model according to our tests or for which no data was available. In these maps we can see that there is indeed a certain correlation between geographical position and location on the $<b,a>$ plane. For example, countries in East Asia are in general far from Zipf and close to disorder, while countries in South America tend to be closer to ideal Zipf's law. \\

\noindent The Euclidean distance on the $<b,a>$ plane also provides us a tool for quantifying the proximity of the internal population distributions within the SAU's of any given countries or regions. We use this distance to measure the similarity of countries according to the internal arrangement of its population. By using the complete linkage clustering method, we get a measure of the cross-correlation between population distributions. Countries exhibiting high cross-correlations are countries with very similar patterns of where people live. Countries in which people spread similarly may have common challenges and perspectives concerning urban planning, population control, sustainability, etc \cite{bettencourt10}. See Fig.(\ref{maps}) for a taxonomic tree of the 108 countries exhibiting good DGBD by this measure. This tree can be divided for any level of correlations into separated families of population-pattern-like countries. By taking the $1/2$ decorrelation point as a cut off, we see that five families of kindred countries arise. Furthermore, closeness of countries in this parameter space shows that they have similar administrative division systems, pointing to akin ways in which these countries divide their land.\\

\section*{The split-merge model}

Suppose we look at a delimited territory, a country for example, where population is spread in communities, cities, towns, villages, etc. The population distribution of this aggregations may be well described by a Pareto, a lognormal distribution or some other function, depending on the specific processes that drove the population growth and dispersion in the region. What happens when a politician decides to create artificial boundaries, dividing the territory in well separated administrative units? We call this units ``municipalities'' during the following discussion. Certainly we are now dealing with a different kind of object and we do not know a priory if the distributions of population in towns and municipalities are the same. It could happen that a big metropolitan area is disaggregated into several municipalities, or that two different villages are grouped together in a common municipality. We \-si\-mu\-la\-ted computationally one particular option for this mechanism by means of what we call the split-merge process: we start with a sample $\bar{X_0}$ of $N_0$ observations following some initial probability distribution $f_0$. Then we picked the two largest values $X_1$ and $X_2$ of the sample and split each of them into two new values, $X_1\longrightarrow p_1X_1,\, (1-p_1)X_1$ and $X_2\longrightarrow p_2X_2,\, (1-p_2)X_2$ where $p_1$ and $p_2$ are random numbers on the interval $(0,1)$. With this, we simulated a process in which two large cities are split into two municipalities each. We randomly chose $3\%$ of the remaining observations and paired them with an equal proportion of randomly selected values. With this  we simulated the action in which $6\%$ of the villages are merged together into new, larger municipalities. With the merged and the split values and the remaining observations we constructed the new sample $\bar{X_1}$ of size $N_1$ following distribution $f_1$. We repeated this process for $n$ iterations; each step represents a system of municipalities in the process of being merged and split in a somewhat arbitrary manner. We wonder what is the distribution of the municipalities populations after several iterations of this process. There are many options about how to proceed with this mechanism; for the present work we chose a particular version of the split-merge process to illustrate the idea.\\

\begin{figure}[!h]
\epsfig{file=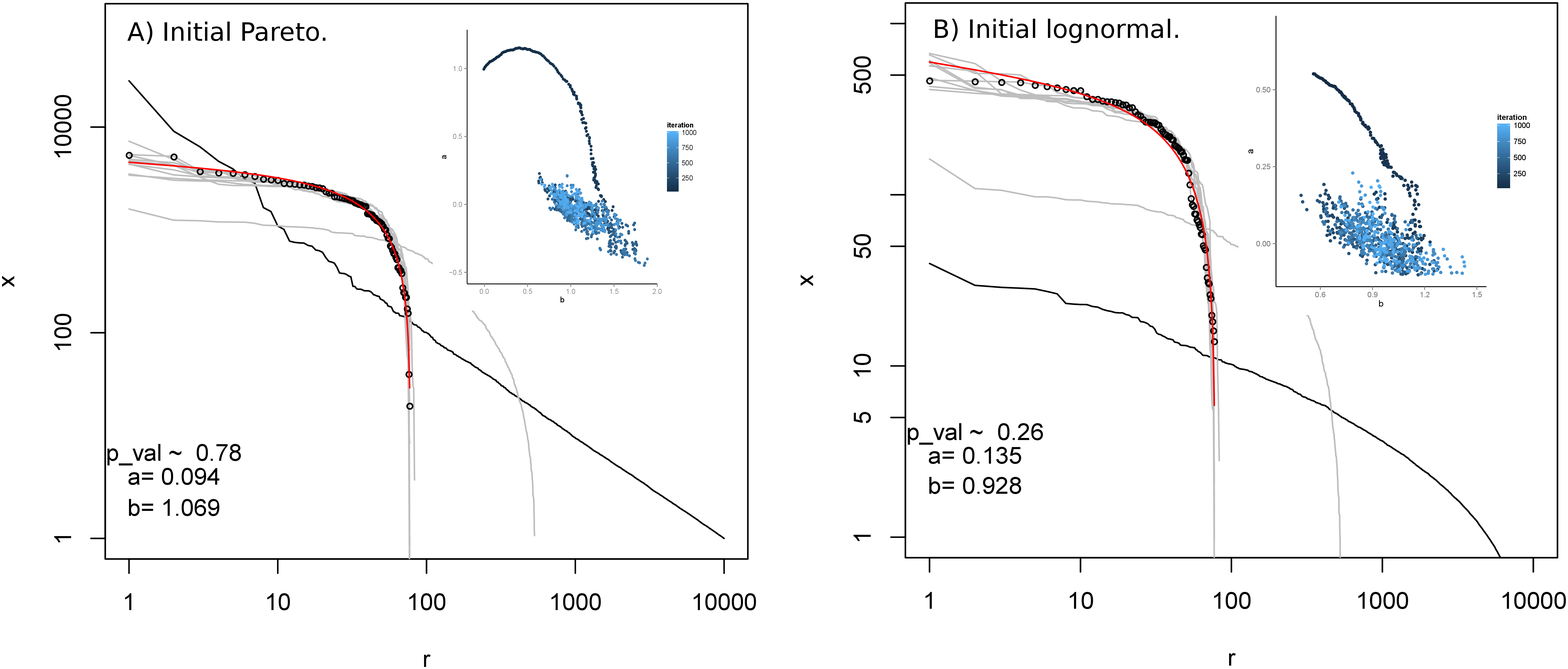,width=0.9\linewidth}
\caption{{\bf Realizations of the split-merge process.} We see in black the initial rank-size function, in gray a sample of intermediate distributions, in circles the final distribution after 1000 iterations and in red the fitted DGBD to the final sample for initial A) Pareto distribution and B) lognormal distribution. On each plot we display p-value of the DGBD hypothesis for the final sample and fitted parameters. Inserted on each panel we show the fitted $b$ and $a$ parameters for each iteration.}

\label{split-merge}
\end{figure}

\noindent We present results for two split-merge process realizations, one with Pareto and one with lognormal as initial distributions. We chose those because they are commonly accepted models of city population distribution. In each process we started with $N_0=10000$; for the initial Pareto we took shape and scale parameters equal to 1, while for the initial lognormal we took location parameter $\mu=0$ and scale parameter $\sigma=1$. We performed $n=1000$ iterations of each process. We were interested in the intermediate rank-size distributions; we fitted this rank-size distribution for each step with the DGBD function, evaluated its performance and analized the temporal evolution of the fitted parameters $b$ and $a$. Fig.(\ref{split-merge}) displays our results: in each frame we see the rank-size  plot for the initial sample, for a selection of 10 intermediate samples (one every 100 iterations) and for the final sample after 1000 iterations. For the final sample we also show the fitted DGBD, as well as the fitted $b$ and $a$ parameters and the estimated p-value of the DGBD hypothesis. To calculate this p-value we used the test described in the previous section. Finally, we show the temporal evolution of the parameters in each realization. \\

\noindent The first thing to observe is that in neither case can we reject the DGBD, according to the estimated p-values. Apparently, DGBD is indeed an adequate model for describing the distributions in this kind of process. Notice how the parameters follow a well defined trajectory on the $<b,a>$ plane at first (dark blue), but begin to move somewhat erratically as times passes (light blue). We also notice how almost invariably $b>a$ for large times. As we mentioned in the previous section the region $a<b$ represents countries whose internal population distribution is somewhere between power law and lognormal. We speculate that these distributions break as the territory is artificially divided into municipalities, leading to new distributions where $b>a$. The fact that sub-samples or aggregations of zipfian sets significantly deviate from Zipf's law has already been observed \cite{cristelli}. Here we have a new result in this direction: a perfect zipfian or lognormal set, describing population in natural agglomerations, transforms into a different kind of set of different kinds of objects, artificial municipalities with arbitrary borders, for which the initial distribution no longer holds. In this study we simulated a very specific split-merge process, whose specific details may show relevant and need a more comprehensive study; nevertheless  we see that this process and the DGBD function provide adequate tools for understanding formation and evolution of administrative divisions.\\

\begin{figure}[!h]
\epsfig{file=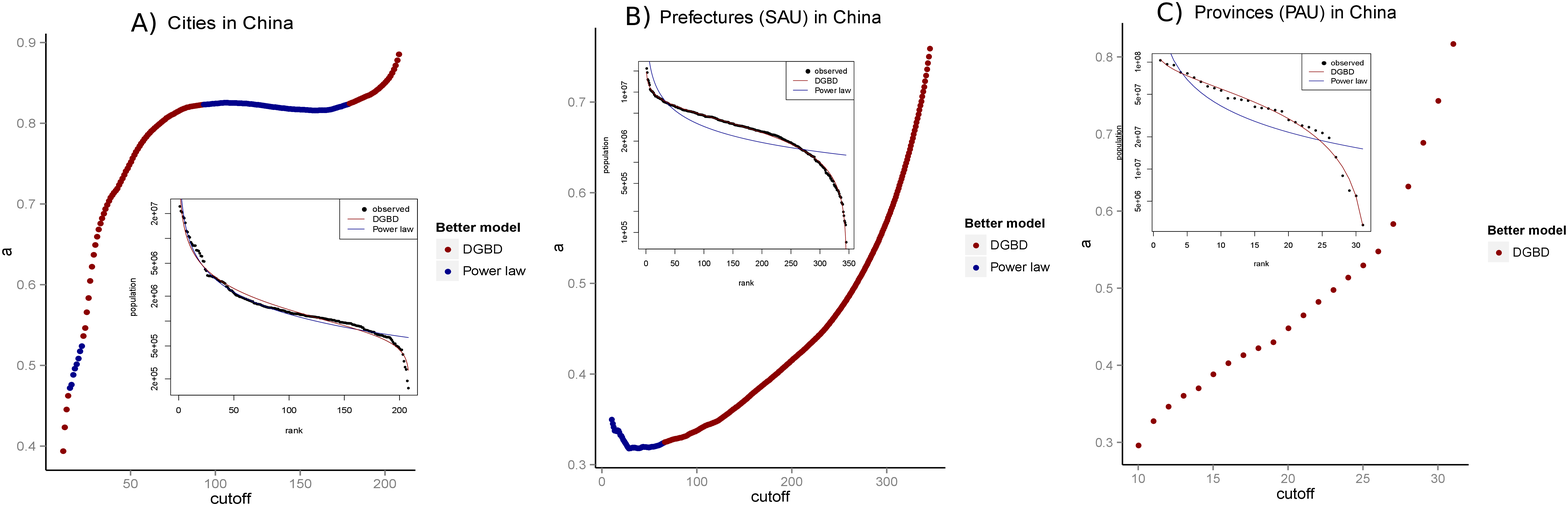,width=0.9\linewidth}
\caption{{\bf Effects of truncation in China population.} Estimated exponent of power law vs. sample size (taking the ``cutoff'' largest units in the sample). According to Akaike information criterion, points marked in red and blue are better described by DGBD and power law respectively. We also see rank-size plot for the whole samples for the distribution population in China of its A) cities, B) prefectures and C) provinces.
}
\label{cutoff}
\end{figure}

\section*{Administrative divisions against natural cities}

\noindent In order to further comprehend the differences between cities and administrative divisions we compared population distribution for cities, primary (PAU) and secondary administrative divisions in a country from our sample. We chose China for this analysis for being the most populated country in the world. City population data was taken from \url{https://en.wikipedia.org/wiki/List_of_cities_in_China_by_population_and_built-up_area} (consulted on June 2016). As we already mentioned, it has been observed that power law fits this kind of data well for the upper tail (big cities), but fails when smaller towns come into consideration, so a cut off is usually to be introduced \cite{peng,giesen}. We took sub-samples of the 10, 11,..., 208 largest cities, fitted each of them to a power law $x(r)=\frac{C}{x^a}$ and estimated the exponent $a$. Fig.(\ref{cutoff}A) shows the number of cities in the sample (the cut off) against the estimated exponent. We see that the parameter is very sensitive to the threshold value, so the decision of where to make it should be taken with extreme care. This truncation may also introduce deviations from power law; for each sub-sample, we tested the goodness of fit of the power law model against the DGBD by the Akaike information criterion. We show in red those sub-samples for whom DGBD is better than power law and in blue those for which power law is better, according to this criterion. We also show the rank-size plot for the whole sample of cities and its respective DGBD and power law fits. With these results, we speculate that truncation produces deviations from power law which are well modelled by DGBD. Aside from truncation, what happens with territorial division comes into play? \\

\noindent The same analysis was performed for the prefectures (SAU) population in China. In Fig.(\ref{cutoff}B) it is possible to see again a high sensibility of the exponent in power law to the cut off value. Except for a few number of small sub-samples, DGBD is a better model than power law in almost every occasion. Now we are not only \-dea\-ling with truncation, but with a process of city splitting as well, as there are natural cities divided into several administration units (for instance, the city of Shanghai is subdivided in 16 SAU's). The splitting mechanism also introduces deviations from power law. Finally we considered provinces (PAU) populations, see Fig.(\ref{cutoff}C). This time DGBD performs better for all sub-samples. When we consider PAU data, the effect of merging becomes visible, because there are some sets of cities grouped into the same unit (for example, the province of Anhui has 12 large cities). This merging operation also causes discrepancies between the observed distribution and power law.  As we see in the rank-size plot, DGBD is a better model for PAU population.\\

\noindent In summary, DGBD is a better model than power law for describing administrative unit population.
Although there is an initial correspondence between cities and administrative divisions, this correspondence ceases to hold when large cities are divided into several units (split) and when separate units are grouped into larger ones (merge), yielding divisions with more than one large city or town. Cities and administrative units are clearly different objects, whose population are described by different kinds of data sets and in most cases they follow different distributions.\\

\section*{Discussion and Conclusions}

\noindent In this work we investigated a seldom discussed subject: formation, evolution and population distribution of administrative units for different countries and territories around the world. Because administrative divisions are artificially defined, serving diverse political and administrative purposes, and are constantly evolving, a high degree of heterogeneity and diversity was to be expected. We tested the commonly used power law against the two parameter DGBD rank-size function for describing the population distribution of second level administrative units. We used a re-sampling approach to the Kolmogorov-Smirnov goodness of fit test to evaluate the fitting of power law and DGBD to the data and the Akaike Information Criterion to compare the performance of these two models. From our population data set, we concluded that DGBD is a good model in $~73.5\%$ of cases, there is only one case that exhibit a good power law (Mayotte, see table \ref{allcountries}), representing $0.7\%$ of the total, and in $25.8\%$ of cases neither is a good model. However, within those countries where both models are rejected by the K-S test, DGBD is better than power law in most of them, just in 1 of 38 cases (Rwanda, see table \ref{allcountries}) power law is better according to AIC. These results, together with visual inspection of Fig.(\ref{150plots}), support the idea that DGBD is an adequate choice when fitting administrative unit population data. As an additional analyzis, we tested the performance of DGBD and power law as statistical models to fit world population by country, which can be thought as a zero-level administrative unit. This is a case in which neither power law nor DGBD are accepted by the K-S test, but DGBD is better according to the AIC criterion.\\

\noindent DGBD allowed us to propose a metric to characterize and compare internal population distribution between different countries and territories. When fitting the data to DGBD, if $b<a$ it is likely that the distribution is somewhere between Pareto and lognormal, as it is for cities and metropolitan areas,  while $b>a$ calls for a different distribution. We derived some analytical relationships between DGBD and uniform distribution, delta distribution, power law and suggested a link to lognormal distribution. This indicates that DGBD is a very flexible function, capable of giving a good representation of a large number of data sets. In consequence, it is a good candidate for characterizing and comparing population distributions in different parts of the world, since they could be following different dynamics, making it difficult to propose a more universal model.\\

\noindent Cities, urban areas and towns are often considered to be organic, natural agglomerations of people living together, while administrative divisions arise from a combination of this self-organized clusters from one side, local and central governments establishing  boundaries from the other. Formation of administrative units is a process driven by controlling agents as well as a certain degree of self organization, showing a high degree of complexity. The ways in which different countries divide their territory obeys disparate purposes but is not completely arbitrary: in general, there is at least one city or town in each administrative division (the capital), large cities are often split into several distinct units, et cetera. \\

\noindent We proposed the split-merge process to simulate the action of bureaucrats and politicians partitioning the territory of a country in  administrative units. Because some cities are divided into distinct units and some units have more than one city, the correspondence between cities and administrative divisions breaks, so we do not have the same population distribution for these two types of  data. Computational simulations show how initial power laws and lognormal distributions of cities evolve into other kind of probabilistic laws for municipalities as the territory gets partitioned; distributions in latter stages are well represented by DGBD with $b>a$. With our numerical evidence, we conjecture that DGBD rises form the split-merge process. Further analysis focusing on the details of the process and a possible analytical derivation of DGBD are still needed.\\

\noindent In addition to the prior discussion, with these results we extend the range of applicability of the DGBD function, contributing to the documentation of its sometimes called  ``universality''. Furthermore, we use for the first time K-S tests and bootstrap methods in this context, which provide a measure of its goodness of fit, and use the Akaike information criterion to evaluate its performance against other models. Our results also rise the question of the performance of  DGBD in describing populations in more ``natural'' urban agglomerations.\\

\noindent In conclusion, DGBD is a better statistical model than power law for fitting administrative unit population data in most of our samples. Deviations from power laws and lognormal distributions arise  as a consequence of a dividing and grouping mechanism. The split-merge process can satisfactorily mimic these mechanisms. The local administrative units approach, the DGBD function and the split-merge model extend the research on population distribution and the undeniable role of artificial boundary settings.

\newpage

\renewcommand\thefigure{A\arabic{figure}}
\renewcommand\thetable{A\arabic{table}}

\section*{Appendix}

\setcounter{figure}{0}
\setcounter{table}{0}

\begin{figure}[!h]
\centering
\epsfig{file=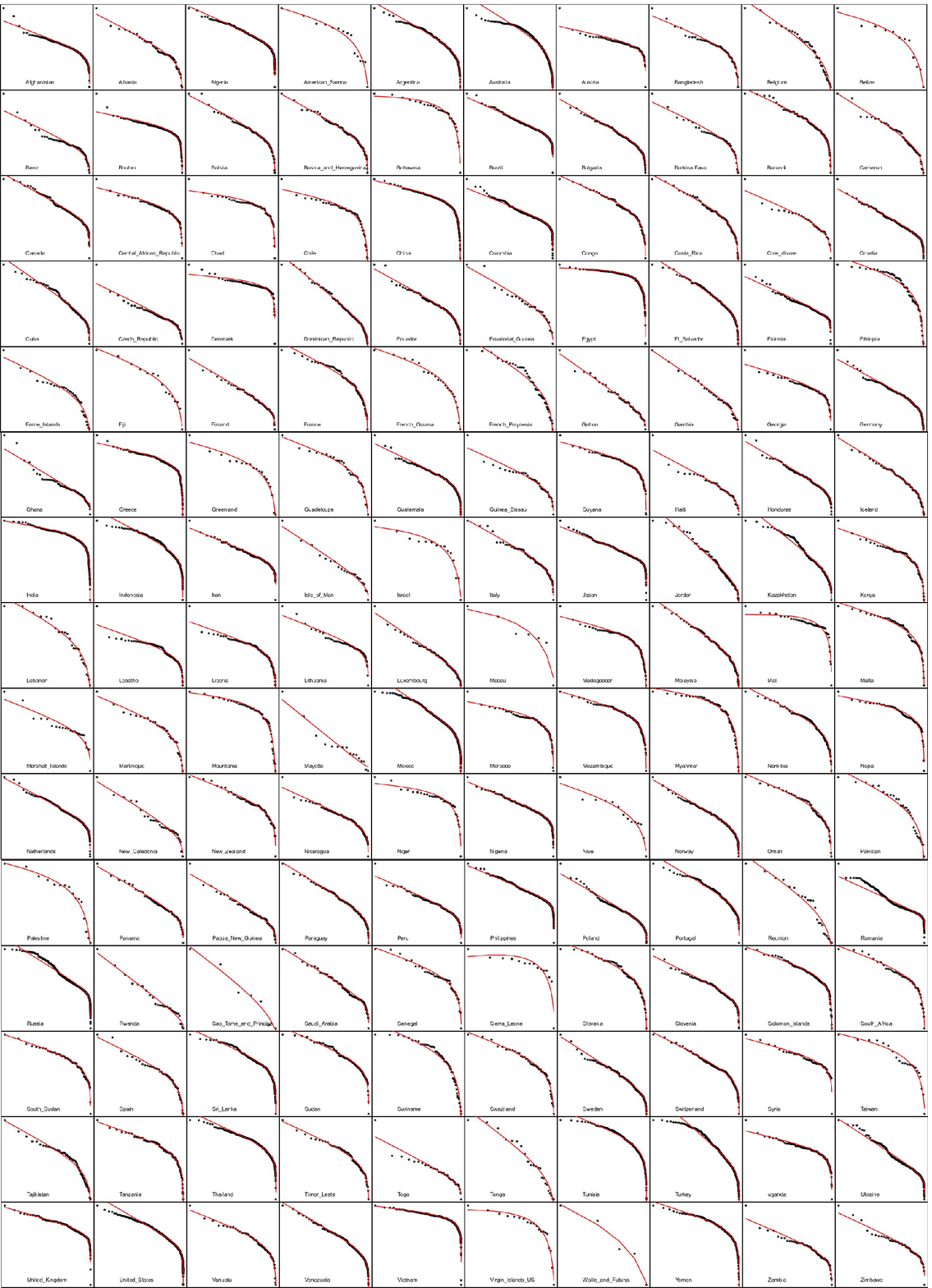,width=0.9\linewidth}
\caption{{\bf Ranked SAU population in log-log scale for the 150 country data base}.  Dots are actual population, red lines are DGBD fits for each data set.
}
\label{150plots}
\end{figure}

Fig.(\ref{150plots}) shows ranked SAU population in logarithmic scale for our whole data base, composed of 150 countries and territories. In the table we show the results of our analysis for the set of 150 countries. For each country, the table displays the fitted $b$ and $a$ parameters, the coefficient of determination of linear regression, number of SAU's $N$, estimated $p$-value with the bootstrap-K-S approach and logarithmic difference of Akaike's Information Criterion, AIC$=log(\mbox{AIC}_{\mbox{\tiny{DGBD}}})-log(\mbox{AIC}_{\mbox{\tiny{power law}}})$.

\scriptsize

\begin{center}

\tabcolsep=0.05cm
\begin{longtable}{|l|cccccc|l|cccccc|}

\caption{Fitted $b$ and $a$ parameters, $R^2$ of lin. regression, sample size (number of SAU's) $N$,  p-val estimation for K-S test and AIC criterion for our samples of countries. AIC indicates $log(\mbox{AIC}_{\mbox{\tiny{DGBD}}})-log(\mbox{AIC}_{\mbox{\tiny{power law}}})$.}\\

\hline
\multicolumn{1}{|c}{\textbf{{\scriptsize country}}} & \multicolumn{1}{c}{\textbf{{\scriptsize b}}} & 
\multicolumn{1}{c}{\textbf{{\scriptsize a}}} & \multicolumn{1}{c}{\textbf{{\scriptsize R}}{\scriptsize $^2$}} & \multicolumn{1}{c}{\textbf{{\scriptsize N}}} &
\multicolumn{1}{c}{\textbf{{\scriptsize p-val}}} & \multicolumn{1}{c|}{\textbf{{\scriptsize AIC}}} &
\multicolumn{1}{|c}{\textbf{{\scriptsize country}}} & \multicolumn{1}{c}{\textbf{{\scriptsize b}}} & 
\multicolumn{1}{c}{\textbf{{\scriptsize a}}} & \multicolumn{1}{c}{\textbf{{\scriptsize R}}{\scriptsize $^2$}} & \multicolumn{1}{c}{\textbf{{\scriptsize N}}} &
\multicolumn{1}{c}{\textbf{{\scriptsize p-val}}} & \multicolumn{1}{c|}{\textbf{{\scriptsize AIC}}} \\
\hline
\endfirsthead	

\multicolumn{14}{c}
{{\bfseries \tablename\ \thetable{} -- continued from previous page}} \\
\hline
\multicolumn{1}{|c}{\textbf{{\scriptsize country}}} & \multicolumn{1}{c}{\textbf{{\scriptsize b}}} & 
\multicolumn{1}{c}{\textbf{{\scriptsize a}}} & \multicolumn{1}{c}{\textbf{{\scriptsize R}}{\scriptsize $^2$}} & \multicolumn{1}{c}{\textbf{{\scriptsize N}}} &
\multicolumn{1}{c}{\textbf{{\scriptsize p-val}}} & \multicolumn{1}{c|}{\textbf{{\scriptsize AIC}}} &
\multicolumn{1}{|c}{\textbf{{\scriptsize country}}} & \multicolumn{1}{c}{\textbf{{\scriptsize b}}} & 
\multicolumn{1}{c}{\textbf{{\scriptsize a}}} & \multicolumn{1}{c}{\textbf{{\scriptsize R}}{\scriptsize $^2$}} & \multicolumn{1}{c}{\textbf{{\scriptsize N}}} &
\multicolumn{1}{c}{\textbf{{\scriptsize p-val}}} & \multicolumn{1}{c|}{\textbf{{\scriptsize AIC}}} \\
\hline
\endhead

\hline \multicolumn{14}{|r|}{{Continued on next page}} \\ \hline
\endfoot

\hline \hline
\endlastfoot			

Afghanistan                & 0.47 & 0.5  & 0.99 & 382  & 0.18   & -455.37  & Macau                    & 1.67  & 0.16  & 0.91 & 7    & 0.01   & 0.84     \\
Albania                    & 0.5  & 0.68 & 0.98 & 61   & 0.19   & -55.1    & Madagascar               & 0.61  & 0.25  & 0.98 & 111  & 0.24   & -155.42  \\
Algeria                    & 0.46 & 0.58 & 1    & 1541 & 0.88   & -3068.41 & Malaysia                 & 0.41  & 0.72  & 0.99 & 144  & 0.41   & -183.16  \\
American Samoa            & 1.82 & 0.72 & 0.95 & 15   & 0.22   & -7.69    & Mali                     & 1.27  & -0.02 & 0.91 & 50   & 0.03   & -40.61   \\
Argentina                  & 0.7  & 0.87 & 0.99 & 511  & 0.2    & -823.66  & Malta                    & 0.72  & 0.35  & 0.98 & 68   & 0.01   & -91.97   \\
Australia                  & 0.96 & 0.89 & 0.97 & 654  & 0      & -641.16  & Marshall Islands        & 0.89  & 1.02  & 0.91 & 25   & 0      & -4.23    \\
Austria                    & 0.49 & 0.35 & 0.9  & 95   & 0.05   & -43.46   & Martinique               & 0.76  & 0.61  & 0.98 & 34   & 0.4    & -36.41   \\
Bangladesh                 & 0.36 & 0.37 & 0.98 & 64   & 0.21   & -62.84   & Mauritania               & 1.08  & 0.14  & 0.97 & 55   & 0.51   & -67.41   \\
Belgium                    & 0.42 & 0.61 & 0.97 & 43   & 0.2    & -24.13   & Mayotte                  & -0.03 & 0.8   & 0.94 & 16   & 0.48   & 4.51     \\
Belize                     & 0.73 & 0.23 & 0.97 & 12   & 0.12   & -6.97    & Mexico                   & 0.79  & 0.93  & 0.99 & 2456 & 0      & -3762.12 \\
Benin                      & 0.18 & 0.38 & 0.94 & 77   & 0.01   & -19.45   & Morocco                  & 0.8   & 0.26  & 0.98 & 75   & 0.04   & -95.22   \\
Bhutan                     & 0.72 & 0.3  & 0.97 & 262  & 0.04   & -302.71  & Mozambique               & 0.64  & 0.3   & 0.99 & 148  & 0.45   & -258.41  \\
Bolivia                    & 0.57 & 0.82 & 0.99 & 112  & 0.92   & -140.6   & Myanmar                  & 0.93  & 0.16  & 0.98 & 63   & 0.31   & -94.98   \\
Bosnia and Herzegovina   & 0.29 & 0.44 & 0.99 & 105  & 0.64   & -150.06  & Namibia                  & 0.34  & 0.34  & 0.99 & 121  & 0.12   & -166.95  \\
Botswana                   & 1.72 & 0.03 & 0.97 & 28   & 0.38   & -33.61   & Nepal                    & 0.81  & 0.25  & 0.97 & 75   & 0.08   & -83.06   \\
Brazil                     & 0.38 & 0.66 & 0.99 & 556  & 0.49   & -629.25  & Netherlands              & 0.35  & 0.6   & 0.99 & 504  & 0.78   & -705.13  \\
Bulgaria                   & 0.4  & 0.75 & 1    & 262  & 0.87   & -424.2   & New Caledonia           & 0.36  & 0.95  & 0.97 & 33   & 0.1    & -8.11    \\
Burkina Faso              & 0.32 & 0.4  & 0.97 & 45   & 0.24   & -29.62   & New Zealand             & 0.67  & 0.62  & 0.98 & 74   & 0.68   & -74.5    \\
Burundi                    & 0.24 & 0.23 & 0.99 & 129  & 0.54   & -217.52  & Nicaragua                & 0.42  & 0.64  & 0.98 & 153  & 0.14   & -144.48  \\
Cameroon                    & 0.31 & 0.54 & 0.98 & 58   & 0.46   & -44.25   & Niger                    & 0.73  & 0.11  & 0.96 & 37   & 0.21   & -36.48   \\
Canada                     & 0.49 & 0.87 & 0.99 & 293  & 0.32   & -329.85  & Nigeria                  & 0.25  & 0.28  & 1    & 775  & 0.75   & -1694.98 \\
Central A. Republic & 0.79 & 0.44 & 0.97 & 72   & 0.32   & -67.95   & Niue                     & 0.67  & 0.44  & 0.95 & 14   & 0.15   & -3.99    \\
Chad                       & 0.7  & 0.17 & 0.95 & 62   & 0.02   & -56.57   & Norway                   & 0.44  & 0.78  & 1    & 431  & 0.38   & -713.23  \\
Chile                      & 1.43 & 0.45 & 0.97 & 54   & 0.06   & -63.42   & Oman                     & 0.83  & 0.61  & 0.99 & 61   & 0.2    & -83.2    \\
China                      & 0.76 & 0.25 & 1    & 345  & 0.51   & -810.67  & Pakistan                 & 0.66  & 0.51  & 0.95 & 30   & 0.05   & -17.27   \\
Colombia                   & 0.51 & 0.7  & 0.99 & 1057 & 0.01   & -1183.06 & Palestine                & 0.8   & 0.22  & 0.96 & 16   & 0.27   & -11.47   \\
Congo                      & 0.49 & 0.43 & 0.99 & 100  & 0.82   & -135.53  & Panama                   & 0.43  & 0.95  & 0.99 & 76   & 0.38   & -77.53   \\
Costa Rica                & 0.46 & 0.52 & 0.99 & 81   & 0.75   & -119.18  & Papua New Guinea       & 0.12  & 0.32  & 0.98 & 87   & 0.05   & -52.71   \\
Cote d'Ivore               & 0.36 & 0.51 & 0.95 & 33   & 0.37   & -12.09   & Paraguay                 & 0.39  & 0.77  & 1    & 224  & 0.59   & -372.36  \\
Croatia                    & 0.31 & 0.77 & 1    & 556  & 0.17   & -705.73  & Peru                     & 0.46  & 0.69  & 0.99 & 194  & 0.67   & -201.04  \\
Cuba                       & 0.24 & 0.58 & 0.99 & 168  & 0.38   & -134.79  & Philippines              & 0.44  & 0.57  & 0.99 & 1634 & 0.19   & -2559.73 \\
Czech Republic            & 0.15 & 0.42 & 0.94 & 77   & 0.07   & -14.45   & Poland                   & 0.17  & 0.46  & 0.99 & 379  & 0.7    & -350.97  \\
Denmark                    & 0.83 & 0.18 & 0.84 & 99   & 0      & -49.92   & Portugal                 & 0.47  & 0.8   & 0.98 & 308  & 0      & -285.39  \\
Dominican Republic        & 0.3  & 0.97 & 1    & 155  & 0.73   & -157.19  & Reunion                  & 0.5   & 0.79  & 0.95 & 24   & 0.02   & -5.78    \\
Ecuador                    & 0.49 & 0.82 & 0.99 & 216  & 0.39   & -251.93  & Romania                  & 0.22  & 0.62  & 0.97 & 2951 & 0      & -1273.87 \\
Equatorial Guinea         & 0.44 & 0.79 & 0.97 & 30   & 0.28   & -11.06   & Russia                   & 0.32  & 0.65  & 0.98 & 2581 & 0      & -2172.22 \\
Egypt                      & 1.4  & 0.1  & 0.99 & 367  & 0.61   & -687.37  & Rwanda                   & 0.02  & 0.15  & 0.98 & 30   & 0.04   & 1.96     \\
El Salvador               & 0.5  & 0.76 & 1    & 262  & 0.87   & -432.23  & Sao Tome and Principe & 0.34  & 0.94  & 0.96 & 7    & 0.25   & 3.42     \\
Estonia                    & 0.3  & 0.78 & 0.96 & 241  & 0.04   & -95.11   & Saudi Arabia            & 0.39  & 0.98  & 0.99 & 118  & 0.01   & -85.43   \\
Ethiopia                   & 0.93 & 0.38 & 0.98 & 66   & 0.03   & -78.56   & Senegal                  & 0.6   & 0.4   & 0.98 & 45   & 0.06   & -51      \\
Faroe Islands             & 0.85 & 0.87 & 0.96 & 34   & 0.37   & -19.39   & Sierra Leone            & 1.39  & -0.26 & 0.85 & 15   & 0.13   & -5.16    \\
Fiji                       & 1.01 & 0.7  & 0.98 & 15   & 0.18   & -11.32   & Slovakia                 & 0.41  & 0.26  & 0.99 & 79   & 0.17   & -117.57  \\
Finland                    & 0.28 & 0.8  & 0.99 & 69   & 0.91   & -43.7    & Slovenia                 & 0.46  & 0.66  & 0.99 & 210  & 0.53   & -263.93  \\
France                     & 0.45 & 0.42 & 0.98 & 96   & 0.12   & -107.51  & Solomon Islands         & 0.62  & 0.35  & 0.99 & 183  & 0.08   & -283.63  \\
French Guiana             & 1.42 & 0.69 & 0.99 & 22   & 0.54   & -27.63   & South Africa            & 0.63  & 0.39  & 0.99 & 52   & 0.75   & -72.7    \\
French Polynesia          & 0.82 & 0.88 & 0.97 & 49   & 0.01   & -31.78   & South Sudan             & 0.44  & 0.3   & 0.98 & 79   & 0.17   & -93.72   \\
Gabon                      & 0.25 & 1.11 & 0.99 & 48   & 0.98   & -20.67   & Spain                    & 0.55  & 0.58  & 0.98 & 52   & 0.74   & -54.25   \\
Gambia                     & 0.31 & 0.9  & 0.99 & 37   & 0.5    & -24.52   & Sri Lanka               & 0.46  & 0.37  & 0.99 & 331  & 0.12   & -581.19  \\
Georgia                    & 0.58 & 0.49 & 0.95 & 66   & 0.07   & -44.94   & Sudan                    & 0.45  & 0.42  & 0.98 & 131  & 0.56   & -150.25  \\
Germany                    & 0.29 & 0.44 & 0.99 & 402  & 0.43   & -523.46  & Suriname                 & 1.08  & 0.52  & 0.96 & 62   & 0.08   & -58.23   \\
Ghana                      & 0.08 & 0.47 & 0.95 & 110  & 0.11   & -7.05    & Swaziland                & 0.35  & 0.21  & 0.98 & 55   & 0.4    & -69.74   \\
Greece                     & 1.11 & 0.38 & 0.99 & 326  & 0.01   & -572.3   & Sweden                   & 0.31  & 0.69  & 0.99 & 289  & 0.03   & -333.33  \\
Greenland                  & 1    & 0.39 & 0.96 & 19   & 0.06   & -13.52   & Switzerland              & 0.49  & 0.56  & 0.99 & 181  & 0.03   & -291.06  \\
Guadeloupe                 & 0.72 & 0.45 & 0.98 & 32   & 0.32   & -35.21   & Syria                    & 0.66  & 0.5   & 0.91 & 61   & 0.06   & -25.96   \\
Guatemala                  & 0.44 & 0.55 & 0.99 & 331  & 0.28   & -484.16  & Taiwan                   & 1.29  & 0.42  & 0.96 & 22   & 0.13   & -17.8    \\
Guinea Bissau             & 0.38 & 0.59 & 0.95 & 39   & 0.5    & -14.28   & Tajikistan               & 0.5   & 0.54  & 0.97 & 75   & 0.01   & -58.83   \\
Guyana                     & 0.97 & 0.48 & 0.98 & 117  & 0.74   & -163.06  & Tanzania                 & 0.47  & 0.27  & 0.99 & 129  & 0.33   & -246.05  \\
Haiti                      & 0.24 & 0.63 & 0.95 & 42   & 0.23   & -6.34    & Thailand                 & 0.41  & 0.39  & 0.99 & 926  & 0      & -1362.43 \\
Honduras                   & 0.29 & 0.69 & 0.99 & 282  & 0.77   & -326.14  & Timor Leste             & 0.44  & 0.43  & 0.99 & 65   & 0.54   & -85.28   \\
Iceland                    & 0.65 & 1.17 & 0.99 & 79   & 0.96   & -97.43   & Togo                     & 0.18  & 0.58  & 0.92 & 35   & 0.45   & -0.24    \\
India                      & 0.94 & 0.22 & 0.99 & 638  & 0.11   & -1270.03 & Tonga                    & 0.51  & 0.9   & 0.97 & 23   & 0.56   & -7.35    \\
Indonesia                  & 0.58 & 0.6  & 0.99 & 497  & 0      & -623.93  & Tunisia                  & 0.43  & 0.26  & 0.99 & 263  & 0.39   & -482.35  \\
Iran                       & 0.37 & 0.6  & 0.98 & 252  & 0.07   & -218.17  & Turkey                   & 0.5   & 0.86  & 0.99 & 923  & 0.01   & -902.32  \\
Isle of Man              & 0.22 & 0.91 & 0.98 & 24   & 0.65   & -1.7     & Uganda                   & 0.47  & 0.26  & 0.98 & 160  & 0.16   & -200.82  \\
Israel                     & 1.04 & 0.12 & 0.96 & 15   & 0.04   & -10.35   & Ukraine                  & 0.21  & 0.59  & 0.99 & 678  & 0.01   & -501.9   \\
Italy                      & 0.32 & 0.54 & 0.99 & 110  & 0.66   & -121.37  & United Kingdom          & 0.39  & 0.32  & 0.96 & 406  & 0      & -379.09  \\
Japan                      & 0.44 & 0.6  & 0.99 & 1180 & 0      & -1443.85 & United States           & 0.7   & 0.92  & 1    & 3143 & 0.05   & -5487.94 \\
Jordan                     & 0.31 & 1.07 & 0.98 & 89   & 0.54   & -39.15   & Vanuatu                  & 0.66  & 0.55  & 0.99 & 62   & 0.5    & -76.95   \\
Kazakhstan                 & 0.25 & 0.6  & 0.98 & 200  & 0.51   & -135.78  & Venezuela                & 0.51  & 0.75  & 1    & 336  & 0.12   & -601.21  \\
Kenya                      & 0.51 & 0.31 & 0.98 & 70   & 0.04   & -89.2    & Vietnam                  & 0.65  & 0.17  & 0.97 & 661  & 0.48   & -828.92  \\
Lebanon                    & 0.62 & 0.56 & 0.97 & 26   & 0.16   & -18.63   & Virgin Islands US      & 1.79  & -0.03 & 0.97 & 20   & 0.2    & -23.04   \\
Lesotho                    & 0.38 & 0.38 & 0.95 & 129  & 0      & -86.26   & Wallis and Futuna      & 0.32  & 0.22  & 0.97 & 5    & 0.03   & 1.92     \\
Liberia                    & 0.73 & 0.56 & 0.98 & 136  & 0.08   & -143.22  & Yemen                    & 0.69  & 0.33  & 1    & 333  & 0.46   & -709.93  \\
Lithuania                  & 0.42 & 0.55 & 0.94 & 60   & 0.01   & -23.37   & Zambia                   & 0.33  & 0.46  & 0.97 & 74   & 0.34   & -47.68   \\
Luxembourg                 & 0.22 & 0.77 & 0.99 & 105  & 0.43   & -77.5    & Zimbabwe                  & 0.29  & 0.49  & 0.97 & 63   & 0.52   & -33.52

\label{longtable}

\end{longtable}

\end{center}

\normalsize

\end{document}